\pdfoutput=1

\documentclass[prl,twocolumn,showpacs,amsmath,amssymb,floatfix,superscriptaddress]{revtex4}
\usepackage{graphicx}
\usepackage{dcolumn}
\usepackage{bm}
\usepackage{amsmath}
\usepackage{amssymb,xcolor,soul}

\usepackage[normalem]{ulem}
\begin{document}
	
	\title{Magnetoelastic and Magnetostrictive Properties of Co$_2$XAl 
	 Heusler Compounds}
	
	\author{Farzad Mahfouzi}
	\email{Farzad.Mahfouzi@gmail.com}
	\affiliation{Department of Physics and Astronomy, California State University Northridge, California, 91330, USA}
	\author{Gregory P. Carman}
	\affiliation{Department of Mechanical and Aerospace Engineering, University of California, Los Angeles, California 90095, USA}
	\author{Nicholas Kioussis}
	\email{nick.kioussis@csun.edu }
	\affiliation{Department of Physics and Astronomy, California State University Northridge, California, 91330, USA}	
     \date{\today}	
	\begin{abstract}
	We present a comprehensive first principles electronic structure study of 
	the magnetoelastic and magnetostrictive properties in the Co-based Co$_2$XAl (X = V, Ti, Cr, Mn, Fe) full Heusler compounds. In addition to the commonly used total energy approach, we employ torque method to calculate the 
	magnetoelastic tensor elements. We show that the torque based methods are in general computationally more efficient,
	 and allow to unveil the atomic- and orbital-contributions to the magnetoelastic constants in an exact manner, as opposed to 
	 the conventional approaches based on second order perturbation with respect to the spin-orbit coupling. 
	 The magnetostriction constants are in good agreement
	  with available experimental data. 
	 The results reveal that the main contribution to the magnetostriction constants, 
	 $\lambda_{100}$ and $\lambda_{111}$, arises primarily from  the strained-induced modulation of the 
	 $\langle d_{x^2-y^2}|\hat{L}_z|d_{xy}\rangle$  and $\langle d_{z^2}|\hat{L}_x|d_{yz}\rangle$ spin orbit coupling matrix elements, respectively, 
	 of the Co atoms.  	
	\end{abstract}

	\maketitle
	
	\section{Introduction}
	Development of efficient and scalable means to manipulate the magnetic state has been one of the main focuses of scientific researches in the field of condensed	matter physics and material science in the past century. The use of magnetoelastic materials employed in multiferroic heterostructures, offers promising avenue for efficient, scalable and nonvolatile magnetic based memory devices\cite{Roy2011}. Magnetoelasticity is a phenomenon where a deformation of the crystal shape results in a change of magnetic orientation, and vice versa.  In addition to applications in multiferroic based magnetic memory devices, compounds with large magnetoelastic constant are also of great interest in the development of efficient magneto-mechanical actuators\cite{Lacheisserie2005}, magnetic field sensors, strain-mediated miniaturized multiferroic-based antennas and other energy converter devices\cite{Clark1980,Andreev1995,Atulasimha2011}. 
	Therefore, development of a concise and efficient framework to calculate the magnetoelastic constants
	 and understand its microscopic origin is of paramount importance in the search for magnetoelastic materials\cite{Fritsch2012,Wu1997,Turek2007}.
	
	Even though the rare-earth-3$d$ metal compounds, such as Terfenol-D, exhibit the highest magnetostriction values (1500-2000 ppm) at room temperature, their use in industrial applications is hindered by the need of high saturation magnetic field (due to their large magnetocrystalline anisotropy), brittleness, and high material costs\cite{Grossinger2008}.  
	Subsequently, highly magnetostrictive rare-earth-free Fe-based alloys 
	were developed, such as Fe$_{1-x}$Ga$_x$ (Galfenol)\cite{Clark2009,Petculescu2005}
	or Fe$_{1-x}$Al$_x$ (Alfenol)\cite{Clark2008}, which display large strain at moderate field and excellent ductility. 
	In addition, spinel ferrites (CoFe$_2$O$_4$, NiFe$_2$O$_4$) with large magnetostriction\cite{Fritsch2012} and high magnetic ordering temperatures 
	have been recently used in magnetostrictive-piezoelectric composites to enhance the interfacial magnetoelectric coupling\cite{Zheng2004}.
	
Another remarkable class of materials are  the Heusler ternary intermetallic compounds that crystallize in the L2$_1$ structure and have stoichiometric composition of X$_2$YZ (space group Fm$\bar{3}$m),  where X and Y are 
transition metal elements and $Z$ is an element from the $p$-block\cite{Graf2011,Felser2016}. 
They show a wide range of remarkable properties such as half-metallicity\cite{Graf2011}, high Curie temperatures\cite{Wurmehl2006},
giant tunnel magnetoresistance\cite{Wang2010,Liu2012}, magnetic shape memory\cite{Ullakko1996}, 
superconductivity\cite{Wernick1983}, topological Weyl Fermions\cite{Graf2011,Wang2016,Belopolski2019}, and the
anomalous Nernst effect\cite{Sakai2018}.
More specifically, the cobalt-based Heusler compounds such as the Co$_2$XAl (X = Ti, V, Cr, Mn, Fe)  offer an interesting playground for spintronics applications since they have high Curie temperatures and some of them are predicted to be half-metallic ferromagnets\cite{Graf2011,Felser2016}. Nevertheless, their magnetoelastic and magnetostrictive properties remain unexplored both experimentally and theoretically.  

Here, we provide a general framework, where we employ different approaches to calculate the magnetoelastic and magnetostriction tensor elements of Co$_2$XAl (X = V, Ti, Cr, Mn, Fe) full Heusler compounds from first principles electronic structure calculations.
The first one is the well-known approach based on total energy calculations and the other two  are based on the torque and spin-orbital torque methods.
We show that the torque based methods are computationally more efficient and allow for the atomic- and orbital-decomposition of the magnetoelastic constants 
which can in turn elucidate the underlying atomic mechanisms.

	\section{Theoretical Formalism}
	\subsection{Magneto-Crystalline Anisotropy}
	The origin of the magnetocrystalline anisotropy (MCA) energy is the spin-orbit interaction and can be determined, 
	within density-functional theory, from the second-variation method employing the scalar-relativistic eigenfunctions 
	of the valence states\cite{Koelling1977,Jansen1990}.  In first principles electronic structure calculations two approaches 
	are often used to calculate the MCA, namely, the total energy and the torque methods.
	
	{\it Total energy approach--} 
 The total energy, $E(\vec{m})$, is determined for several magnetic orientations described by the unit vector, $\vec{m}$,
 which in turn is fitted to lowest order in the magnetic degrees of freedom, given by,

	\begin{equation}
 	E_{tot}(\vec{m}) = 	E_{tot}^0 +  \sum_{ij}K^{ij}m_im_j. 
 	\label{TotEn}
 \end{equation}			
Here, $K^{ij} $ are the MCA tensor matrix elements and $m_i$'s are the components of the magnetization orientation unit vector, $\vec{m}$. 

As an alternative approach, instead of the total energy one can employ the so-called force theorem\cite{Weinert1985} where the dependence of the electronic energy on the 
magnetization directions can be approximately expressed in terms of the band energies, $E_{band}$, (sum of occupied one-electron eigenvalues), namely, 
	\begin{equation}\label{eq:bandE}
	E_{band}(\vec{m}) =\frac{1}{N_k}\sum_{n\vec{k}}\epsilon_{n\vec{k}}^{\vec{m}}f(\epsilon_{n\vec{k}}^{\vec{m}}-\mu(\vec{m})).
	\end{equation}
Here, $f(x)$ is the Fermi-Dirac distribution function, $N_k$ is the number of $k$-points,
and $\mu(\vec{m})$ is the electronic chemical potential which depends on the 
magnetization direction. 
	
	{\it  Torque Approach--} 
Wang {\it et al.}  proposed\cite{Wang1996} a torque method for the theoretical determination of the MCA energy for systems with uniaxial symmetry, where instead of directly calculating the total energy difference it involves the expectation value of the angular derivative of the SOC Hamiltonian at an angle $\theta$=45$^{\circ}$, 
	\begin{equation}
	T(\theta) = \sum_{n{\bf k}}^{occ} \Big<\Psi_{n{\bf k}}^{SOC}\Big\vert\frac{\partial H^{SOC}}{\partial \theta}\Big\vert
	\Psi_{n{\bf k}}^{SOC}\Big>_{\theta=45^{\circ}}. 
	\end{equation}
Here, $\Psi_{n{\bf k}}^{SOC}$ is the $n$th relativistic eigenvector at {\bf k} point, and $\theta$ is the angle between the magnetization direction and the surface normal. 

The one-electron Kohn-Sham Hamiltonian can be expressed by\cite{Mahfouzi2017,Mahfouzi2018}, 
		\begin{align}
	\hat{H}&=\hat{H}_K(\vec{k})\hat{1}_{2\times 2}+\hat{\Delta}(\vec{k})\vec{m}\cdot\vec{\hat{\sigma}}+\hat{H}_{soc}(\vec{k}),
	\end{align}
	where, the first, second and third terms represent the kinetic, exchange, and SOC contributions, respectively. In a non-orthonormal atomic orbital basis set, the eigen-energies/states are calculated from the generalized eigenvalue problem, $\hat{H}|n\vec{k}\rangle=\epsilon_{n\vec{k}}\hat{\mathcal{O}}|n\vec{k}\rangle=\epsilon_{n\vec{k}}{\mathcal{O}}_{n\vec{k}}|n\vec{k}\rangle$, where $\hat{\mathcal{O}}(\vec{k})$ is the overlap matrix. In this case, the torque is given by,\cite{Mahfouzi2018}
	\begin{align}
	\vec{\tau}_{MCA}=-\vec{m}\times\langle\hat{\Delta}\vec{\hat{\sigma}}\rangle,\label{Trq}
	\end{align}
	 where the equilibrium expectation value is calculated from,
	\begin{align}
	\langle...\rangle&=\frac{1}{N_k}\sum_{n\vec{k}}\langle n\vec{k}|...|n\vec{k}\rangle \frac{f(\epsilon_{n\vec{k}}-\mu_0)}{\mathcal{O}_{n\vec{k}}}.
	\label{2}
	\end{align}
	Unlike total energy method, the torque approach involves a vector for the fitting to the magnetization orientation and also it does not require the calculation of a reference energy, making it computationally more efficient. Furthermore, the torque method can be used to calculate the local (site-resolved)-contribution to the MCA energy, since the exchange splitting, $\hat{\Delta}$, is often a well-defined local quantity.
	
	In this manuscript, instead of the aforementioned torque method we employ a different approach we have recently developed\cite{Mahfouzi2020},  
	based on the canonical forces, $F_{\theta}=\vec{n}\cdot\vec{\tau}=-\langle\frac{\partial\hat{H}}{\partial\theta}\rangle$ and $F_{\phi}=\vec{e}_z\cdot\vec{\tau}=-\langle\frac{\partial\hat{H}}{\partial\phi}\rangle$, where $\theta$ ($\phi$) is the polar (azimuthal) angle,  $\vec{n}=\sin{\phi}\vec{e}_x-\cos{\phi}\vec{e}_y$, and 
	$\vec{e}_z$ is the unit vector along $z$. 
	Applying the unitary operator $\hat{U}=e^{i\theta\vec{n}\cdot\vec{\hat{\sigma}}/2}$ on the Hamiltonian to reorient the exchange splitting term along the z-axis we find	
	\begin{align}
	F_{q}&=2Re\big\langle\hat{U}\frac{\partial \hat{U}^{\dagger}}{\partial q}\hat{H}_{soc}\big\rangle,\ \  q=\theta,\phi .
	\label{RotForce}
	\end{align}
	Using Eq.~\eqref{RotForce} for $q=\theta$,  one can obtain an explicit expression for the MCA induced torque, 
	\begin{align}
	\vec{\tau}_{MCA}=\langle\hat{\xi}\vec{\hat{L}}\times\vec{\hat{\sigma}}\rangle,\label{RotTrq}
	\end{align}
	which we refer it as the ``spin-orbital'' torque approach\cite{Mahfouzi2017,Mahfouzi2020} as opposed to the original torque method given by Eq.~\eqref{Trq}.  It should be pointed out that Eq.~\ref{RotTrq} is exact and no approximation was involved in its derivation.
	
	Eq.~\eqref{RotTrq} can be interpreted as the torque induced by the anisotropic orbital moment accumulation, $\vec{\hat{L}}$, on the spin, $\vec{\hat{\sigma}}$, of the valence electrons. Since the SOC strength, $\hat{\xi}$, is diagonal in the atomic-orbital basis set and a well-defined local quantity, we can use Eq.~\eqref{RotTrq} to decompose the torque on each atom. This decomposition allows in turn to elucidate the atomic origin of 
    the MCA as opposed to the local MCA induced field on each atomic-spin given by Eq.~\eqref{Trq}.
	Therefore, the advantage of using Eqs.~\eqref{RotForce} and ~\eqref{RotTrq} is that they allow  to unveil the 
	underlying origin of the MCA. 
	Employing Eq.~\eqref{RotTrq} the atom- and orbital-contribution to the total torque can be written as, 
	\begin{align}
	\langle \alpha|\vec{\hat{\tau}}^I_{MCA}|\beta\rangle&=\sum_{ss'}\rho_{ss'}^{I,\alpha\beta}\langle I\alpha s|\hat{\xi}\vec{\hat{L}}\times\vec{\hat{\sigma}}|I\beta s'\rangle,\label{Orb_res_Fq}
	\end{align}
	where, $I$ is the atomic index, $\alpha,\beta$ ($s,s'$) are the orbital (spin) indices, and 
	\begin{align}
	\rho_{ss'}^{I,\alpha\beta}&=\frac{1}{N_k}\sum_{n\vec{k}}\langle I\beta s'|n\vec{k}\rangle \frac{f(\epsilon_{n\vec{k}}-\mu_0)}{\mathcal{O}_{n\vec{k}}}\langle n\vec{k}|I\alpha s\rangle,
	\end{align}
	is the density matrix.

	\subsection{Magneto-Elastic Effect}
	Magnetoelastic  coupling is the  interaction  between  the  magnetization and the strain in a magnetic material.
	In the presence of strain, ${\varepsilon_{ij}}$, the modified primitive lattice vectors, $\vec{a}'_i$, are given by $(\vec{a}'_i-\vec{a}_i)\cdot\vec{e}_j=\sum_{k}\vec{a}_{i}\cdot\vec{e}_k\varepsilon_{kj}$, where the  $\vec{e}_j$'s represent unit vectors in Cartesian coordinates.
 To lowest order in the lattice deformation ({\it i.e.} small strain) and magnetization orientation, the total energy per equilibrium volume is given by,
	\begin{align}
	E (\vec{m})=E_0+\frac{1}{2}\sum_{i\le j,k\le l}C^{ij}_{kl} \varepsilon_{ij}\varepsilon_{kl}+\sum_{ij}K^{ij} (\{\varepsilon_{kl}\})m_im_j,
	\label{EqEM}
	\end{align}			
	where, $C_{kl}^{ij}$s are the elastic stiffness constants, often represented by a $6\times 6$ matrix. To linear order in strain, the 
	MCA tensor matrix elements are of the form, 
	$K^{ij}  (\{\varepsilon_{kl}\})=K_0^{ij}+\sum_{k\le l}B^{ij}_{kl}\varepsilon_{kl}$, where the $B^{ij}_{kl}$ denote the magnetoelastic tensor elements. 
	
	The magnetostriction effect, first  identified  in 1842 by James Joule\cite{Joul1842}, is  a  property  of ferromagnetic materials  that  causes them  to  change  their  shape  when  subjected  to  a magnetic  field. In the absence of an external stress, the strain induced on the crystal structure due to the reorientation of the magnetization, can be calculated by setting, $\partial E(\vec{m})/\partial \varepsilon_{kl}=0$, 
	 \begin{align}
	 \varepsilon_{kl}=-\sum_{kl}h^{ij}_{kl}m_{i}m_{j},
	 \label{MagStrain}
	 \end{align}	
	 where $h_{kl}^{ij}=\sum_{k'\le l'}S_{kl}^{k'l'}B_{k'l'}^{ij}$ are the magnetostriction tensor elements and 
	  $S_{kl}^{ij}$ are the elastic compliance constants. Under the applied strain, $\varepsilon_{ij}$, 
	  the relative change of the length of the material, $\delta l/l$ along a direction given by the unit vector $\vec{u}$ can be calculated \cite{Kittel1949,lee} from, $\delta l/l=\sum_{ij}\varepsilon_{ij}u_iu_j$. Using Eq.~\eqref{MagStrain} for the strain, the relative change of the length due to the reorientation of the magnetization can be calculated from, 
	 \begin{align}
	\frac{\delta l}{l}=-\sum_{ijkl}h^{ij}_{kl}u_iu_jm_{k}m_{l}.\label{MagStrict}
	\end{align}	 
	
		Given, that the components of the unit vectors $\vec{u}$ and $\vec{m}$ describing the directions of the relative change of the length 
		and magnetization, respectively, are not independent, the basis set in Eq. (\ref{MagStrict}) consisting of $u_iu_j$ and $m_im_j$ is overcomplete. One approach to resolve this issue is to switch to the spherical Harmonics basis set,\cite{Collins} which is 
		more advantageous, specially, when dealing with ensemble averaging. In the following we use this approach to obtain a general expression for the polycrystalline magnetostriction constant. Using the second order spherical Harmonics we can rewrite Eq.~\eqref{MagStrict} in the form
	\begin{align}
	\frac{\delta l}{l}=\sqrt{\frac{4\pi}{5}}\sum_{p}\lambda^{(0)}_{p}Y_{2,p}(\vec{m})+\frac{4\pi}{5}\sum_{pq}\lambda^{(2)}_{pq}Y_{2,p}(\vec{m})Y_{2,q}(\vec{u}),
	\end{align}	 	
	where the isotropic (volumetric) magnetostriction constant, $\lambda^{(0)}_p$ $(p=1,\ldots,5)$, and anisotropic magnetostriction constants,  
	$\lambda^{(2)}_{pq}$, can be expressed (see Appendix I)  in terms of the $h_{kl}^{ij}$, and 
	$Y_{2,p}$'s are the real spherical harmonics, given by, 
	\begin{align}
	Y_{2,p}(\vec{r})=\sqrt{\frac{15}{4\pi}}\Big(\frac{x^2-y^2}{2},\frac{3z^2-1}{2\sqrt{3}},yz,xz,xy\Big).
	\end{align}	   
	
	For a polycrystalline structure the field-induced relative change  of the length is of the form $\delta l/l=\lambda_sP_2(\vec{m}\cdot\vec{u})$, where $P_2(x)$ denotes the Legendre polynomials of order 2. Therefore, the average magnetostriction constant $\lambda_s$ can be calculated from,	 
	\begin{align}
	\lambda_s=\frac{5}{(4\pi)^2}\iint d\Omega_{\vec{m}}d\Omega_{\vec{u}}\frac{\delta l}{l}P_{2}(\vec{m}\cdot\vec{u})=\frac{1}{5}\sum_{p}\lambda^{(2)}_{pp}.
	\end{align}	 
	For a cubic crystal structure the magnetostriction constant matrix, $\lambda_{pq}^{(2)}$, is diagonal and the magnetic field-induced shape deformation is given by,
	
\begin{align}
\frac{\delta l}{l}=\frac{4\pi}{5}\Big[&\lambda_{[100]}\sum_{p=1,2}Y_{2,p}(\vec{u})Y_{2,p}(\vec{m})~+ \\
&\lambda_{[111]}\sum_{p=3,4,5}Y_{2,p}(\vec{u})Y_{2,p}(\vec{m})\Big].\nonumber
\end{align}  
In this case, for the polycrystalline magnetostriction constant we obtain, $\lambda_s=(2\lambda_{[100]}+3\lambda_{[111]})/5$\cite{Kittel1949}.
	
	\section{Computational Approaches}
	We have employed two {\it ab initio} electronic structure codes to determine the magnetoelastic tensor elements. The first is the
	plane wave Vienna {\it ab initio} simulation package (VASP) \cite{Kresse96a,Kresse96b} where we have employed the total energy 
	approach. The second is the linear combination of atomic orbitals (LCAO) OpenMX package\cite{OzakiPRB2003,OzakiPRB2004,OzakiPRB2005}, where one can employ either one of the four approaches, namely, the total energy, the band energy (Eq.~\ref{eq:bandE}), the torque (Eq.~\ref{Trq}),
	or the ``spin orbital" torque (Eq.~\ref{RotTrq}) approach. Throughout the remaining manuscript all OpenMX results employ the more computationally efficient 
	spin orbital torque approach. 
	
	(1) Structural relaxations were carried out using VASP \cite{Kresse96a,Kresse96b} within the generalized gradient approximation (GGA) as parameterized by Perdew et al.\cite{PBE}(PBE)
	when the largest atomic force is smaller than 0.01 eV \AA$^{-1}$. The pseudopotential and wave functions are treated within the projector-augmented wave (PAW) method \cite{Blochl94,KressePAW}. The plane wave cutoff energy was set to 500 eV and a 18$^3$ k-points mesh was used in the Brillouin Zone (BZ) sampling. 
	Total energy calculations were carried out for 9 different 
	magnetization orientations, $\vec{m}$ = [1,0,0], [0,1,0], [0,0,1], [1,1,0], [1,$\bar{1}$,0], [1,0,1], [1,0,$\bar{1}$], [0,1,1], and [0,1,$\bar{1}$], respectively. 
	 The MCA tensor elements in Eq.~\eqref{EqEM} were then calculated from
\begin{subequations}
	\begin{align}
	K^{zz} &= 0\\
	K^{xx} &= E^{[1,0,0]} - E^{[0,0,1]},\\ 
	K^{yy}&=E^{[0,1,0]}-E^{[0,0,1]},  \\
	K^{xy}&=\frac{E^{[1,1,0]}-E^{[1,\bar{1},0]}}{2}, \\
	K^{yz}&=\frac{E^{[0,1,1]}-E^{[0,1,-1]}}{2},  \\
	K^{xz}&=\frac{E^{[1,0,1]}-E^{[1,0,-1]}}{2}. 
	\end{align}	 
\end{subequations}
	
	(2)  Using the lattice parameters determined from VASP calculations, the tight-binding Hamiltonian, $\hat{H}_{\vec{k}}$ and overlap, $\hat{\mathcal{O}}_{\vec{k}}$ matrices were calculated in the LCAO OpenMX package\cite{OzakiPRB2003,OzakiPRB2004,OzakiPRB2005}.
	We adopted the Troullier-Martins type norm-conserving pseudopotentials\cite{TroullierPRB1991} with partial core correction. 
	We used $24^3$ k-points in the first BZ, and an energy cutoff of 350 Ry for numerical integrations in the real space grid. For the exchange correlation functional the LSDA\cite{CeperleyPRL1980} parameterized by Perdew and Zunger\cite{PerdewPRB1981} was used.
	The MCA tensor elements, $K_{ij}$, are determined 
	via the spin-orbital torque (Eq.~\eqref{RotTrq}) method for 
 three magnetization directions, $\vec{m}=$$[1,0,0],$ $[1,0,1]$, and $[0,1,1]$,
 respectively, from the expressions, 
\begin{subequations}
\begin{align}
\vec{\tau}^{[100]}_{MCA}&=[0,2K^{xz},-2K^{xy}],\\
\vec{\tau}^{[101]}_{MCA}&=[K^{xy}+K^{yz},K^{zz}-K^{xx},-K^{xy}-K^{yz}],\\
\vec{\tau}^{[011]}_{MCA}&=[K^{yy}-K^{zz},-K^{xy}-K^{xz},K^{xy}+K^{xz}].
\end{align}
\end{subequations}	 

The magnetoelastic constant tensor elements, $B_{ij}^{kl}$, are determined from MCA calculations under 12 strain, $\varepsilon_{ij}$, values of,   $\varepsilon_{xx}=\pm \delta_{\varepsilon}$, $\varepsilon_{yy}=\pm \delta_{\varepsilon}$, $\varepsilon_{zz}=\pm \delta_{\varepsilon}$, $\varepsilon_{xy}=\pm \delta_{\varepsilon}$, $\varepsilon_{yz}=\pm \delta_{\varepsilon}$, $\varepsilon_{xz}=\pm \delta_{\varepsilon}$, where, $\delta_{\varepsilon}=0.01$. The magnetoelastic constant tensor elements are then simply given by,	
	\begin{align}\label{eq:eq20}
	B^{ij}_{kl}&=\frac{K^{ij}(\varepsilon_{kl}=\delta_{\varepsilon})-K^{ij}(\varepsilon_{kl}=-\delta_{\varepsilon})}{2\delta_{\varepsilon}}.
	\end{align}	 
	
It should be noted that the symmetry of the crystal structure can significantly reduce the number of independent configurations (induced strain and magnetization directions) required to obtain the magnetoelastic tensor elements. In particular, in cubic systems, only two nonzero independent magnetoelastic constants exist that are referred to as, $B_1=B^{xx}_{xx}=B^{yy}_{yy}=-B^{xx}_{zz}=-B^{yy}_{zz}$ and $B_2=B^{xy}_{xy}=B^{yz}_{yz}=B^{zx}_{zx}$, constants corresponding to the normal and shear induced MCAs, respectively.
	
	\section{Results and Discussion}
\begin{figure} [tbp] 
		{\includegraphics[angle=0,trim={0cm 0cm 0.0cm 0cm},clip,width=0.5\textwidth]{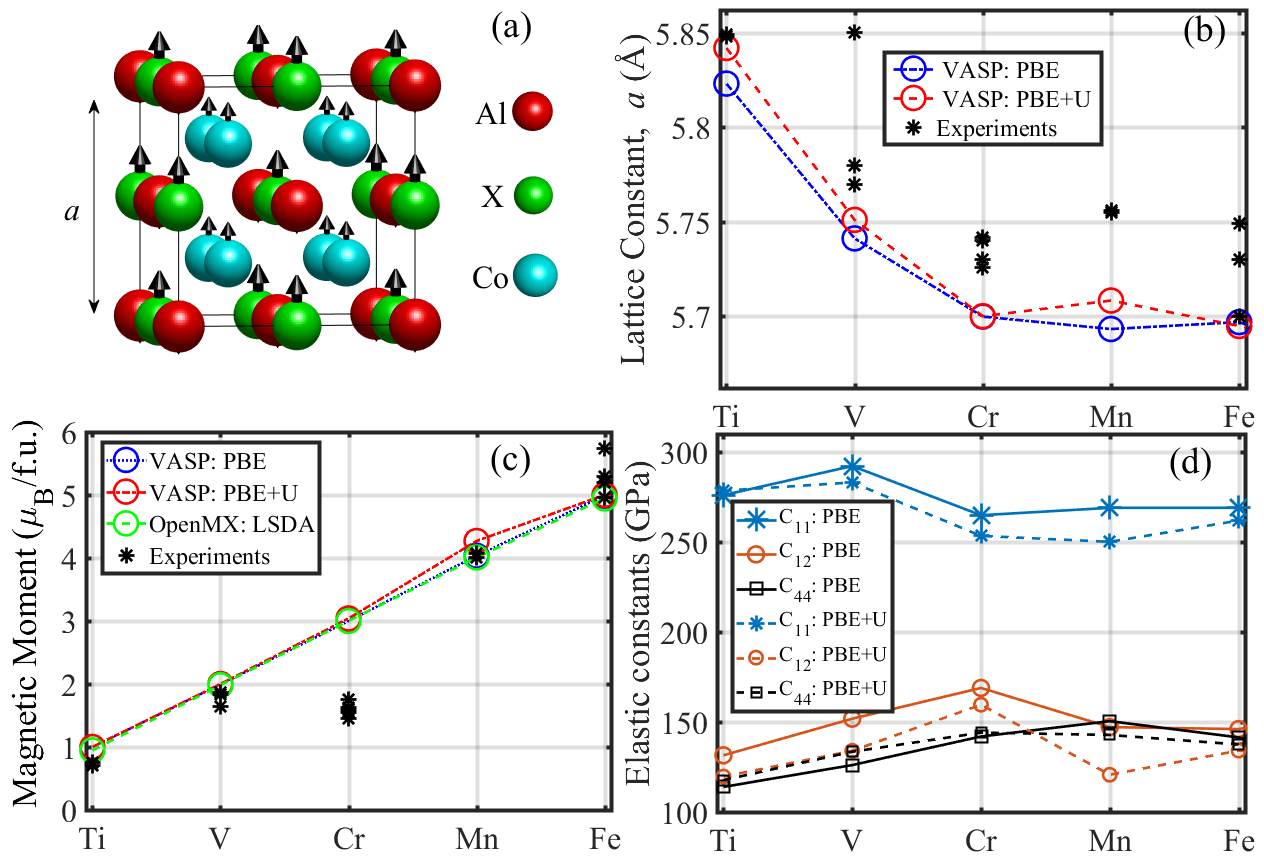}}%
		\caption{(a) L2$_1$ crystal structure of full Heusler compounds. (b) Lattice constants of Co$_2$XAl compounds using PBE exchange correlation functional with (red circles) and without (blue circles) Hubbart U included\cite{Kandpal}. The stars show  experimental data reported in \cite{1.Graf2009,2.Webster,4.Kanomata,5.Carbonari,6.Ziebeck,7_15.BUSCHOW,8.Hakimi,9.Kudryavtsev,10.Kourova,11.Nehla,12.Nehla,13.Webster,14.Umetsu,16.Jain,17.Husain,18.elmers}.  (c) Total magnetic moment per formula unit versus X elements using VASP (blue and red symbols) and OpenMX (green symbols). The experimental results, shown as black stars for X=Ti, V, Cr, Mn and Fe have been reported in \cite{1.Graf2009,2.Webster}, \cite{4.Kanomata,5.Carbonari,6.Ziebeck}, \cite{7_15.BUSCHOW,8.Hakimi,9.Kudryavtsev,10.Kourova,11.Nehla,12.Nehla}, \cite{13.Webster,14.Umetsu} and  \cite{7_15.BUSCHOW,16.Jain,17.Husain,18.elmers}, respectively.(d) Elastic constants, $C_{11}$, $C_{12}$, and $C_{44}$, calculated using PBE (dashed lines)  and PBE+U (solid lines) exchange-correlation functional in VASP.}
		\label{fig:fig1}
	\end{figure}

The Heusler compounds Co$_2$XAl crystallize in the cubic L2$_1$ structure 
(space group Fm$\bar{3}$m) which is shown in the inset of Fig.~\ref{fig:fig1}(a). The Co atoms occupy the Wyckoff position 8c (1/4, 1/4, 1/4), the X and the Al atoms are located at 4a (0, 0, 0) and 4b (1/2, 1/2, 1/2), respectively. As depicted in Fig.~\ref{fig:fig1}(a), this structure consists of four interpenetrating fcc sublattices, two of which are equally occupied by X\cite{Graf2011,Felser2016}. 

The calculated lattice constants shown in Fig.~\ref{fig:fig1}(b) demonstrates a monotonic decrease with increasing atomic number of the X element, consistent with their corresponding atomic radius.  
We have also carried out PBE+U calculations where we used the values of U for the $d$-orbitals of Co and the X elements from Ref. \cite{Kandpal}.  	
	The effect of U on the lattice constants (blue dashed curve in Fig.~\ref{fig:fig1}(b)) shows a slight increase of the lattice constant when compared to the case without U. The results 
	are in good agreement with the experimentally reported data \cite{1.Graf2009,2.Webster,4.Kanomata,5.Carbonari,6.Ziebeck,7_15.BUSCHOW,8.Hakimi,9.Kudryavtsev,10.Kourova,11.Nehla,12.Nehla,13.Webster,14.Umetsu,16.Jain,17.Husain,18.elmers}, denoted by black star symbols in Fig.~\ref{fig:fig1}(b).
Heusler compounds are known for their well behaved magnetic properties in terms of their total number of valence electrons. The total 
magnetic moment per formula unit is shown in Fig.~\ref{fig:fig1}(c) versus the X element (sorted with respect to its atomic number). In agreement with the 
 Slater-Pauling curve\cite{book_Bozorth}, the magnetic moment per formula unit are integer numbers that depend linearly on the number of valence electrons per formula unit, $N_v$, given by, $M_s=N_v-24$ ($M_s=34-N_v$) for X$\le $Fe (X$\ge$ Fe).  Surprisingly, the results are relatively insensitive to the exchange correlation functional (PBE, PBE+U or LSDA) and except for Co$_2$CrAl, the {\it ab initio} results are in relative 
 	good agreement with the experimentally  reported findings in  Refs. \cite{1.Graf2009,2.Webster,4.Kanomata,5.Carbonari,6.Ziebeck,7_15.BUSCHOW,8.Hakimi,9.Kudryavtsev,10.Kourova,11.Nehla,12.Nehla,13.Webster,14.Umetsu,16.Jain,17.Husain,18.elmers}. The slight increase of the magnetic moment in Co$_2$MnAl due to the inclusion of U is in agreement with previous DFT calculations.\cite{Kandpal} The origin of the discrepancy in the case of Co2CrAl is attributed to B2-like disorder and an antiferromagnetic coupling of Cr with its neighbors, leading to ferrimagnetic behavior\cite{Kubler}.

	For cubic crystal structures the elastic energy is given by,
	\begin{align}
	E_{el}&=\frac{1}{2}C_{11}(\varepsilon^2_{xx}+\varepsilon^2_{yy}+\varepsilon^2_{zz})+\frac{1}{2}C_{44}(\varepsilon^2_{xy}+\varepsilon^2_{yz}+\varepsilon^2_{xz})\nonumber\\
	&+C_{12}(\varepsilon_{xx}\varepsilon_{yy}+\varepsilon_{yy}\varepsilon_{zz}+\varepsilon_{xx}\varepsilon_{zz}), 
	\end{align}	 
	where the subscripts in $C_{ij}$ correspond to the Voigt notation ($[1,2,3,4,5,6]\equiv [xx,yy,zz,yz,xz,xy]$). In Fig.~\ref{fig:fig1}(d) we present the calculated (using VASP) elastic constants, $C_{11}$, 
	 C$_{12}$ and $C_{44}$, versus X elements. 
The results are in good agreement with previous first principles electronic structure calculations.\cite{Felser2019} The solid (dashed) lines in Fig.~\ref{fig:fig1}(d) correspond to the DFT calculations without (with) the Hubbard U term. The inclusion of U results in an overall decrease of the $C_{11}$ and $C_{12}$ elastic constants and a small change of $C_{44}$. Elastic stability of a compound requires that all eigenvalues of the 6$\times$6 elastic matrix be positive. For a cubic crystal structure the eigenvalues are, $C_{44}$, $C_{11}+2C_{12}$ and $C_{11}-C_{12}$, corresponding to shear, bulk and tetragonal shear moduli, respectively. The results for the elastic constants presented in Fig.~\ref{fig:fig1}(b) demonstrates that all compounds are stable under any elastic deformation.

\begin{figure} [tbp] 
	{\includegraphics[angle=0,trim={0.0cm 2cm 0.0cm 2.5cm},clip,width=0.5\textwidth]{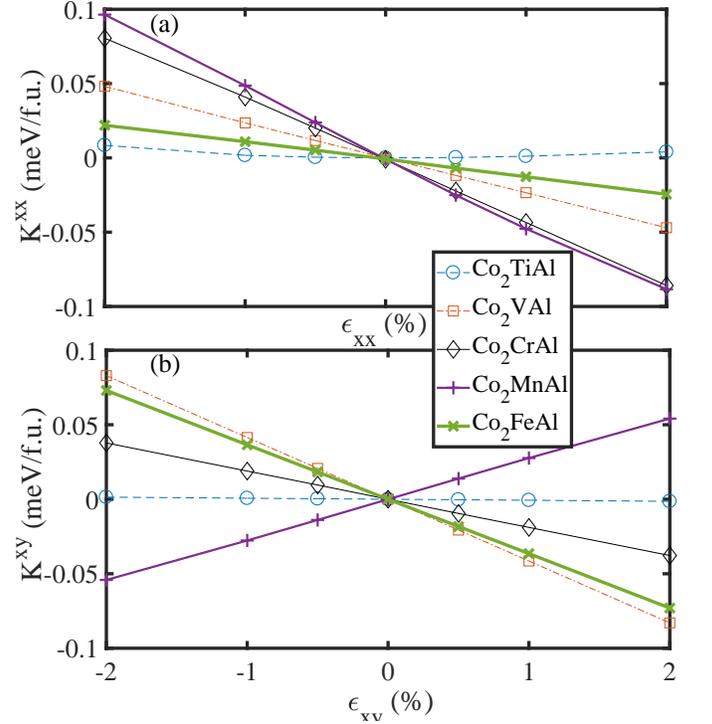}}%
	\caption{ Strain dependence of magnetocrystalline anisotropy coefficients, $K^{xx}$, and $K^{xy}$, calculated from the "spin orbital" torque approach under $\varepsilon_{xx}$ and $\varepsilon_{xy}$ strain, respectively, for the Co$_2$XAl (X =  Ti, V, Cr, Mn, Fe) family.}
	\label{fig:fig2}
\end{figure}

The magnetoelastic energy for a cubic crystal structure is given by,\cite{Kittel1949}
	\begin{align}
	E_{me}&=B_1 (\varepsilon_{xx}m_x^2+\varepsilon_{yy}m_y^2+\varepsilon_{zz}m_z^2)\nonumber\\
	&+B_2 (\varepsilon_{xy}m_xm_y+\varepsilon_{yz}m_ym_z+\varepsilon_{xz}m_xm_z).
	\label{EqEM1}
	\end{align}	 	
	 
	 Fig.~\ref{fig:fig2} shows the magnetocrystalline anisotropy tensor matrix elements, $K^{xx}$ and $K^{xy}$, as a function of strain, $\varepsilon_{xx}$ and $\varepsilon_{xy}$, respectively, for the
	 Co$_2$XAl Heusler compounds, 
	 using the spin-orbital torque approach with the OpenMX DFT package.
	 As expected, the strain dependence is linear within the range of -2\% to +2\%, suggesting that two strain values, as implemented in Eq.~\eqref{eq:eq20}, are sufficient to calculate the magnetoelastic coefficients accurately. Note that $dK^{xx}/d\varepsilon_{xx}<0$ for all compounds. On the other hand, the variation of 
	 $dK^{xy}/d\varepsilon_{xy}$ across the series is non-monotonic and is 
	 discussed in detail below.

\begin{figure} [tbp]
	{\includegraphics[angle=0,trim={0cm 2.9cm 0.0cm 2.cm},clip,width=0.5\textwidth]{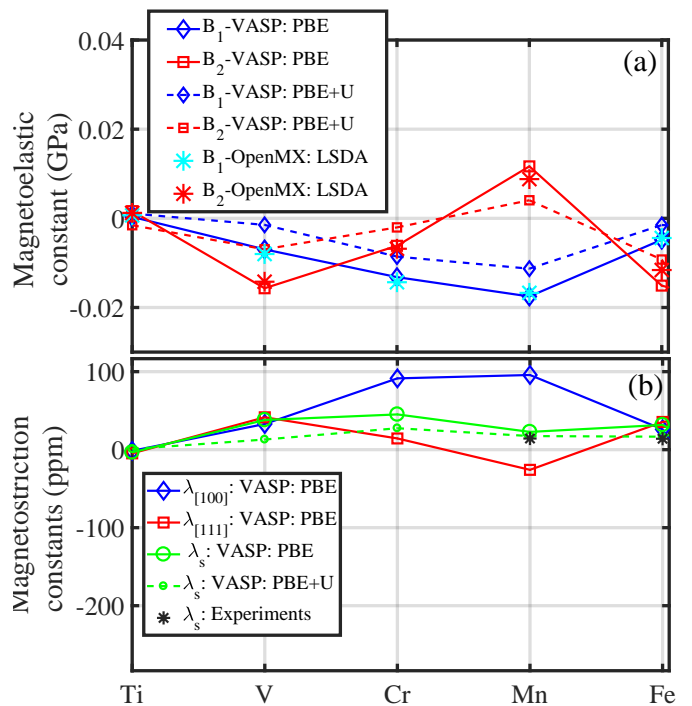}}%
	\caption{(a) Magnetoelastic constants, $B_1$ (blue symbols) and $B_2$ (red symbols), versus X elements in Co$_2$XAl Heusler compounds, calculated using the VASP with PBE exchange correlation (solid lines) and PBE+U (dashed lines). We have also included results of calculation using OpenMX with LSDA exchange correlation functional shown as star symbols. 
			(b) Magnetostriction constants, $\lambda_{[100]}$ and $\lambda_{[001]}$, (using VASP with PBE) and the average magnetostriction  for polycrystalline systems, $\lambda_s$, versus X elements. The dashed line corresponds to the polycrystalline magnetostriction calculated using VASP with PBE+U. For comparison we also show the experimental values
		(black stars) for Co$_2$MnAl (Ref.~\cite{Qiu2008}) and Co$_2$FeAl (Ref.~\cite{Gueye2014}) at room temperature.}
	\label{fig:fig3}
\end{figure}

	 Fig.~\ref{fig:fig3}(a) displays the magnetoelastic constants $B_1$ and $B_2$ versus the X element, shown as blue and red symbols respectively. The solid (dashed) lines in Fig.~\ref{fig:fig3}(a) are the results of VASP calculations using PBE without (with) the U term, while the stars are calculated using OpenMX with the LSDA exchange-correlation functional. We find an overall good agreement between the results of the two different {\it ab initio} packages. The effect of U is to reduce both magnetoelastic constants by a factor of two.
	 Fig.~\ref{fig:fig3}(a) shows that the magnetoelastic constant, $B_1$, is negative for all members of the Co$_2$XAl family independent of the exchange correlation functionals and {ignoring the effect of Hubbard U it} ranges from 
	 around -20 MPa to 0 MPa, comparable to the corresponding range for the spinel ferrites CoFe$_2$O$_4$ and NiFe$_2$O$_4$\cite{Fritsch2012}.
	 The magnetoelastic coupling constants $B_2$ range from about -15 MPa to +10 MPa, which are higher by an order of magnitude compared to the corresponding values for the spinel ferrites.  
	 In Fig.~\ref{fig:fig3}(b) we show the magnetostriction constants, $\lambda_{[100]}$ and $\lambda_{[111]}$, and the average magnetostriction constant, $\lambda_s$, suitable for polycrystalline systems, versus the X element. The polycrystalline magnetostriction constant using PBE+U (dashed green curves) is approximately 50\% lower than the corresponding values without U (solid green curves). Since, the difference between the magnetoelastic constants obtained from VASP and OpenMX is small, we show in Fig.~\ref{fig:fig3}(b) only the magnetostriction constants calculated from VASP. For comparison we also display the available experimental values of $\lambda_s$ for Co$_2$MnAl,\cite{Qiu2008} and  Co$_2$FeAl,\cite{Gueye2014}. 
	 Overall, the DFT+U results are in better agreement with the experimentally reported room-temperature values. It should be noted that, since thermal spin and phonon fluctuations are not taken into account in the DFT calculations, one should not expect a very good agreement between the theoretical results and the reported experimental values at room temperature.

\begin{figure} [tbp]
		{\includegraphics[angle=0,trim={0.0cm 4.0cm 0.0cm 2.0cm},clip,width=0.5\textwidth]{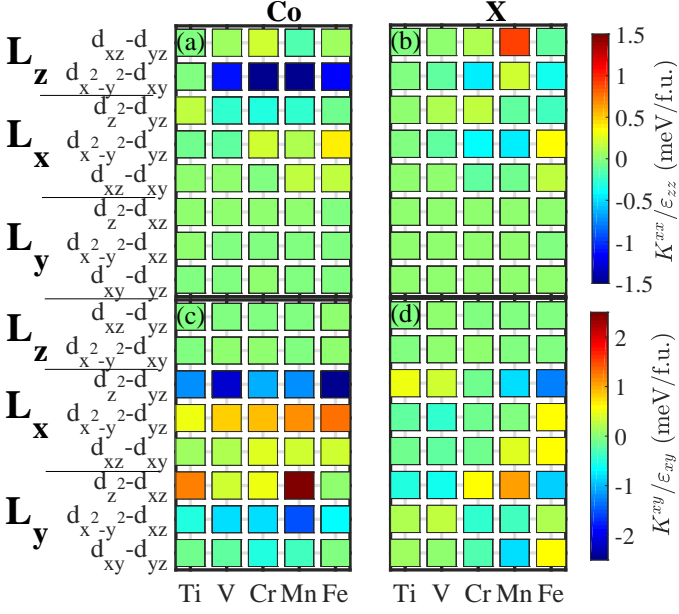}}%
		\caption{ (a,c) Co and (b,d) X projected atomic orbital resolved contributions to strain induced MCA, $K^{xx}/\varepsilon_{zz}$ and  $K^{xy}/\varepsilon_{xy}$ shown as top (a,b) and bottom (c,d) figures, respectively. The left-hand ordinate shows the 
			nonzero matrix elements of the three components of the orbital angular momentum operators, $\hat{L}_y$, $\hat{L}_x$ and $\hat{L}_z$. }
		\label{fig:fig4}
	\end{figure}

	To understand the underlying origin of the magnetoelastic properties across the series we have used Eq.~\eqref{Orb_res_Fq}  employed in the OpenMX DFT package to resolve the total torque into its atomic and orbital contributions. 
	In Fig.~\ref{fig:fig4} top (bottom) we show the orbital and atomic contribution to the magnetoelastic constants, $K^{xx}$ ($K^{xy}$) 
	versus X-elements. 
	The MCA constants originate primarily from the Co and X elements, 
	shown on the left and right panels, respectively. 
	On the left-hand ordinate in Fig.~\ref{fig:fig4} we display the 
	nonzero matrix elements of the three components of the orbital angular momentum operators, $\hat{L}_y$, $\hat{L}_x$ and $\hat{L}_z$.
	
	For a cubic crystal structure subject to strain along z, the nonzero MCA constant, $K_{xx}=K_{yy}$, is given by,	
	\begin{align}\label{EqKxx}
	K^{xx}=-\vec{\tau}_{MCA}^{[101]}\cdot\vec{e}_y=\big\langle\hat{\xi}(\hat{L}_x\hat{\sigma}_z-\hat{L}_z\hat{\sigma}_x)\big\rangle^{[101]},
	\end{align}	 
	where the first and second terms correspond to the in-plane (xy-plane) and out-of-plane (z-axis) contribution of the strain-induced orbital moment accumulation, respectively. 
	This is consistent with Figs.~\ref{fig:fig4}(a,b), where, except for the case of X=Co, the magnetoelastic constant, $B_1$ is dominated by the contribution of the strain-induced $\hat{L}_z$ orbital moment accumulation of the Co atoms. 
	The $\langle d_{x^2-y^2}|\hat{L}_z|d_{xy}\rangle$ contribution to $B_1$ can be further decomposed into the spin-diagonal and 
	spin-off-diagonal components, where, according to the second order perturbation approach, the former (later) yields
	positive (negative) contributions to the uniaxial MCA. Under a tensile strain along $z$ 
	we find a significant reduction of $\langle d_{x^2-y^2}|\hat{L}_z|d_{xy}\rangle$  
	resulting in a negative sign for $B_1$.
	 
	
	Similarly, using the spin-orbital torque expression, the strain-induced MCA under biaxial $\varepsilon_{xy}$ strain the  
	magnetoelastic constant, $K^{xy}$, is given by the expression, 
		\begin{align}\label{EqKxy}
	K^{xy}=-\frac{1}{2}\vec{\tau}_{MCA}^{[100]}\cdot\vec{e}_z=-\frac{1}{2}\big\langle\hat{\xi}(\hat{L}_x\hat{\sigma}_y-\hat{L}_y\hat{\sigma}_x)\big\rangle^{[100]}.
	\end{align}	 
	In the rotated frame of reference where the magnetization is along $z$, 
	Eq.~\eqref{EqKxy} shows that the spin-diagonal (-off-diagonal) matrix elements contribute to the 
 orbital moment accumulation along $y$ ($x$). Similar to the $K^{xx}$ magnetoelastic constant, the main contribution to $K^{xy}$ arises from the Co atoms,  
 where the negative sign of $B_2$ is mainly due to the $\langle d_{z^2}|\hat{L}_x|d_{yz}\rangle$ orbital momentum matrix element. 
 The sign reversal of $K^{xy}$ for X=Mn, is due to the relatively large positive contribution to the strain induced-orbital moment 
 accumulation along the $y$-axis.

	


\section{CONCLUSION}
In summary, we have presented a detailed first-principles study of the magnetoelastic and magnetostrictive properties of Co$_2$XAl full Heusler compounds that crystallize in the L2$_1$ structure. We described three computational approaches to calculate the magnetoelastic and magnetostriction tensor matrix elements. The first one is the well-known approach based on total energy calculations. The other two novel approaches,  are based on the  torque\cite{Mahfouzi2018} and spin-orbital torque\cite{Mahfouzi2020} approaches, respectively.
The latter two are computationally more efficient and allow the atomic- and orbital-decomposition of the magnetoelastic constants 
which can in turn elucidate the underlying atomic mechanisms.  
In addition, a general approach was presented to determine the average magnetostriction constants, suitable 
for polycrystalline systems,  in terms of the magnetostriction tensor matrix elements.
The results of the different computational approaches, using both the VASP and OpenMX packages, 
agree well and they are also in good agreement with available experimental data. 

\section{ACKNOWLEDGMENTS}
The work is supported by NSF ERC-Translational Applications of Nanoscale Multiferroic Systems (TANMS)- Grant No. 1160504.
We would like to thank N. Jones and K. B. Hathaway for useful discussions. 
	
\appendix

\section{Appendix A}	\label{AppA}
The isotropic (volumetric) magnetostriction constants, $\lambda_{p}^{(0)},$ and anisotropic magnetostriction constants, $\lambda_{pq}^{(2)}$, can be expressed in terms of the magnetostriction tensor elements, $h_{ij}^{kl}$, 
	\begin{subequations}
		\begin{align}
		\lambda_{1}^{(0)}&=\frac{1}{9}\sum_i(2h_{ii}^{zz}-h_{ii}^{xx}-h_{ii}^{yy})\\
		\lambda_{2}^{(0)}&=\frac{1}{3\sqrt{3}}\sum_i(h_{ii}^{xx}-h_{ii}^{yy})
		\end{align}	
	\end{subequations}
	\begin{subequations}
	\begin{align}
	\lambda_{11}^{(2)}&=\frac{1}{9}(4h_{zz}^{zz}+h_{xx}^{xx}+h_{yy}^{yy}+h_{xx}^{yy}-2h_{zz}^{xx}-2h_{zz}^{yy}\nonumber\\
	&+h_{yy}^{xx}-2h_{xx}^{zz}-2h_{yy}^{zz})\\
	\lambda_{22}^{(2)}&=\frac{1}{3}(h_{xx}^{xx}+h_{yy}^{yy}-h_{xx}^{yy}-h_{yy}^{xx})\\
	\lambda_{12}^{(2)}&=\frac{1}{3\sqrt{3}}(2h_{zz}^{xx}-h_{xx}^{xx}-h_{yy}^{xx}-2h_{zz}^{yy}+h_{xx}^{yy}+h_{yy}^{yy})\\
	\lambda_{21}^{(2)}&=\frac{1}{3\sqrt{3}}(2h_{xx}^{zz}-h_{xx}^{xx}-h_{xx}^{yy}-2h_{yy}^{zz}+h_{yy}^{xx}+h_{yy}^{yy})\\
	\lambda_{1p}^{(2)}&=\frac{2}{3\sqrt{3}}(2h_{zz}^{p}-h_{xx}^{p}-h_{yy}^{p}),\ \ p=yz,xz,xy\\
	\lambda_{2p}^{(2)}&=\frac{2}{3}(h_{xx}^{p}-h_{yy}^{p}),\ \ p=yz,xz,xy\\
	\lambda_{pq}^{(2)}&=\frac{4}{3}h_{p}^{q},\ \ p,q=yz,xz,xy
	\end{align}	
	\end{subequations}
where, we used the following expressions,	
\begin{subequations}
	\begin{align}
	&\int{d\Omega}d_{z^2}(x^2)=-\sqrt{\frac{4\pi}{45}}\\
	&\int{d\Omega}d_{z^2}(y^2)=-\sqrt{\frac{4\pi}{45}}\\
	&\int{d\Omega}d_{z^2}(z^2)=\sqrt{\frac{16\pi}{45}}\\
	&\int{d\Omega}d_{x^2-y^2}(x^2)=\sqrt{\frac{4\pi}{15}}\\
	&\int{d\Omega}d_{x^2-y^2}(y^2)=-\sqrt{\frac{4\pi}{15}}\\
	&\int{d\Omega}d_{x^2-y^2}(z^2)=0.
	\end{align}	 
\end{subequations}

%

	
	
\end{document}